%% file: Master.tex
\documentclass[tightenlines,a4paper,12pt,aps]{revtex4}
\usepackage{equations}
\usepackage{epsfig}
\usepackage{graphicx}

\newenvironment{equation*}{\begin{displaymath}}{\end{displaymath}}
\newcommand{\const}{\mbox{const} }
\newcommand{\Scri}{\mbox{$\cal J$}}

\newcommand{\DIII}{\,{}^{\scriptscriptstyle(3)\!\!\!\:}\nabla}
\newcommand{\DeIII}{\,{}^{\scriptscriptstyle(3)\!\!\!\:}\Delta}
\newcommand{\RIII}{\,{}^{\scriptscriptstyle(3)\!\!\!\:}R}

\newcommand{\RII}{\,{}^{\scriptscriptstyle(1,1)\!\!\!\:}\hat R}

\def\@warning#1{\typeout{LaTeX Warning [l.\the\inputlineno]: #1.}}

\begin{document}


\input Titel
\input Einfuehrung
\input Anfangsdaten
\input Zeitentwicklung
\input Zusammenfassung

\input biblio

\end{document}

%% file: Titel.tex
\title{From Now to Timelike Infinity on a Finite Grid}

\author{Peter H\"ubner}
\email{pth@aei-potsdam.mpg.de}
\affiliation{%
  Max-Planck-Institut f\"ur Gravitationsphysik\\
  Albert-Einstein-Institut\\
  Am M\"uhlenberg 1\\
  D-14476 Golm\\
  FRG}

\begin{abstract}
We use the conformal approach to numerical relativity to evolve
hyperboloidal gravitational wave data without any symmetry assumptions.
Although our grid is finite in space and time, we cover the whole
future of the initial data in our calculation, including future null
and future timelike infinity.
\end{abstract}

\maketitle


%% file: Einfuehrung.tex
%
%
%
\section{Introduction}
\hskip-\parindent{}%
In the articles~\cite{Hu99ht,Hu99as,Hu00nc,HuWXXxx} we presented a
complete code for solving the conformal Einstein equations which will
allow us to study many interesting questions about the global
structure of spacetimes by performing numerical experiments. 
In the present paper we want to demonstrate some of the unique
capabilities of this code by a simple example:
We calculate the time evolution of gravitational wave data without
continuous symmetries.
The data are prescribed on a hyperboloidal, whence
spacelike, initial slice extending to future null infinity.
Starting from this slice, we calculate a conformal spacetime $(M,
g_{ab})$ by solving the conformal time evolution equations.
On the region $\tilde{M}$ of the conformal spacetime on which the
variable $\Omega$, the conformal factor, is positive, the metric
$\tilde{g}_{ab} = \Omega^{-2} g_{ab}$ defines an asymptotically flat solution
to Einstein's vacuum field equations. 
We call $(\tilde{M}, \tilde{g}_{ab})$ the physical spacetime. The boundary
of $\tilde{M}$ in $M$, given by the set $\{\Omega = 0\}$,  represents 
future null infinity, ${\cal J}^+$, and future timelike infinity,
$i^+$.
\\
In our evolution the relevant part of the set $\{\Omega = 0\}$ is
completely covered by a numerical grid. 
Therefore, embedding the physical spacetime into a larger conformal
spacetime implies that the determination of the
gravitational radiation is a well-defined, gauge ambiguity-free
procedure and that we avoid any influence of artificial boundaries.
It also allows very accurate determination of the fall-off behaviour
of physical quantities near the different infinities.
\\
Our data are chosen sufficiently close to Minkowski data, so that the
solution admits a regular point $i^+$ in the conformal extension.
However, they are also far enough away to produce a spacetime
which differs significantly from Minkowski space. 
It has been known theoretically for some time that sufficiently
weak data should admit a regular point $i^+$ at timelike
infinity~\cite{Fr87ot} (cf.\ also~\cite{ChK93TG} for similar results
for weak Cauchy data).
However, it is a quite remarkable fact that this point can be
modelled in a numerical calculation with the precision discussed
below.
\\
In the present paper we shall not attempt to discuss the background of
the conformal approach again.
We refer the reader to~\cite{Hu99ht} and to the recent survey article
by J.~Frauendiener~\cite{Fr00ci}.
In section~\ref{AnfangsdatenAbschnitt} we shall describe the given data.
Then we discuss their evolution
(section~\ref{Zeitentwicklungsabschnitt}) and show, in particular,
that we have indeed covered the initial hypersurface as well as future
null infinity and timelike infinity.
%
%
%

%% file: Anfangsdaten.tex
%
%
%
\section{The initial data}
\label{AnfangsdatenAbschnitt}
\begin{figure}[htbp]
  \begin{center}
    \begin{minipage}[t]{16cm}
      \includegraphics[width=7.75cm]{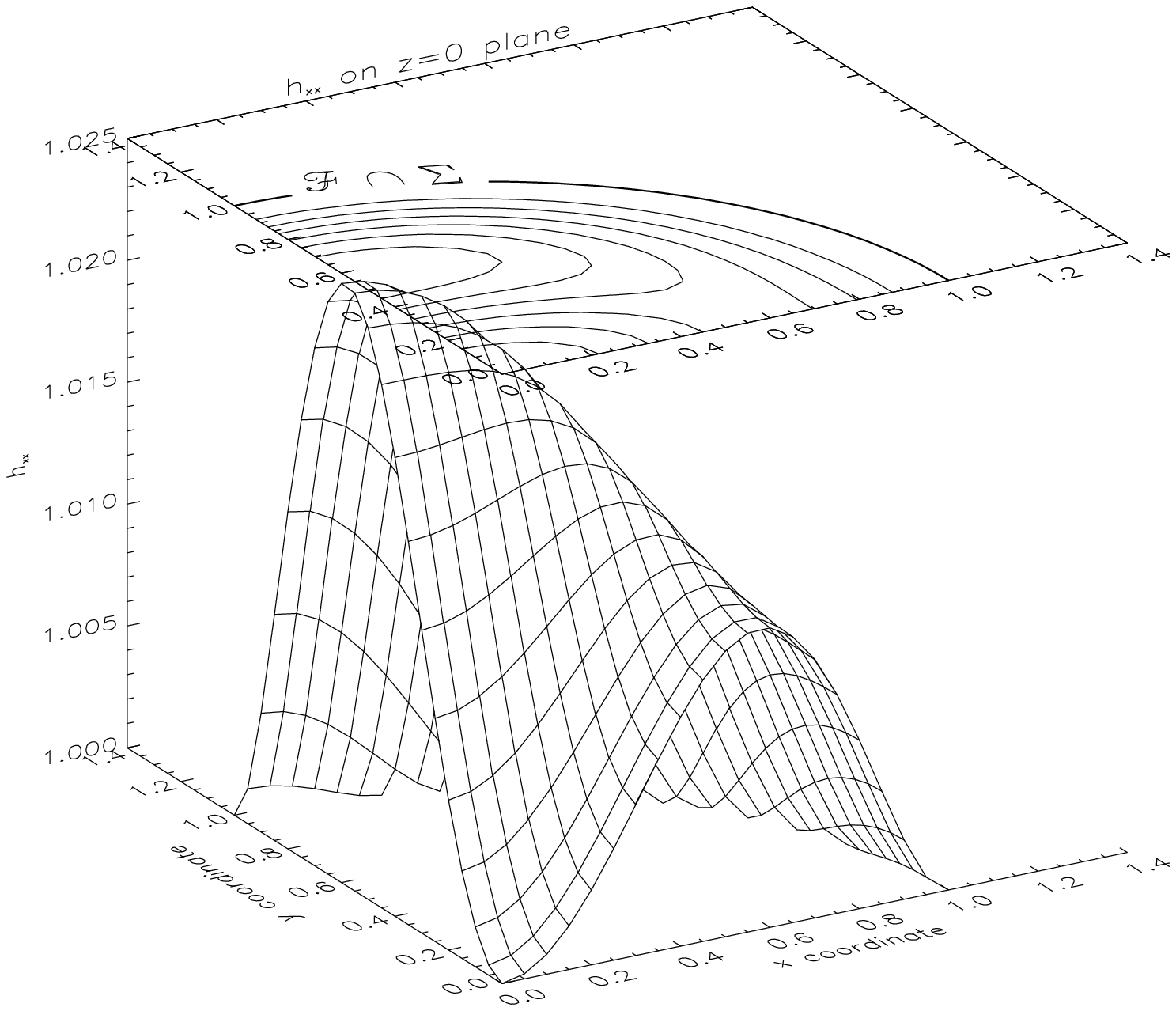}%
      \hspace{.5cm}
      \includegraphics[width=7.75cm]{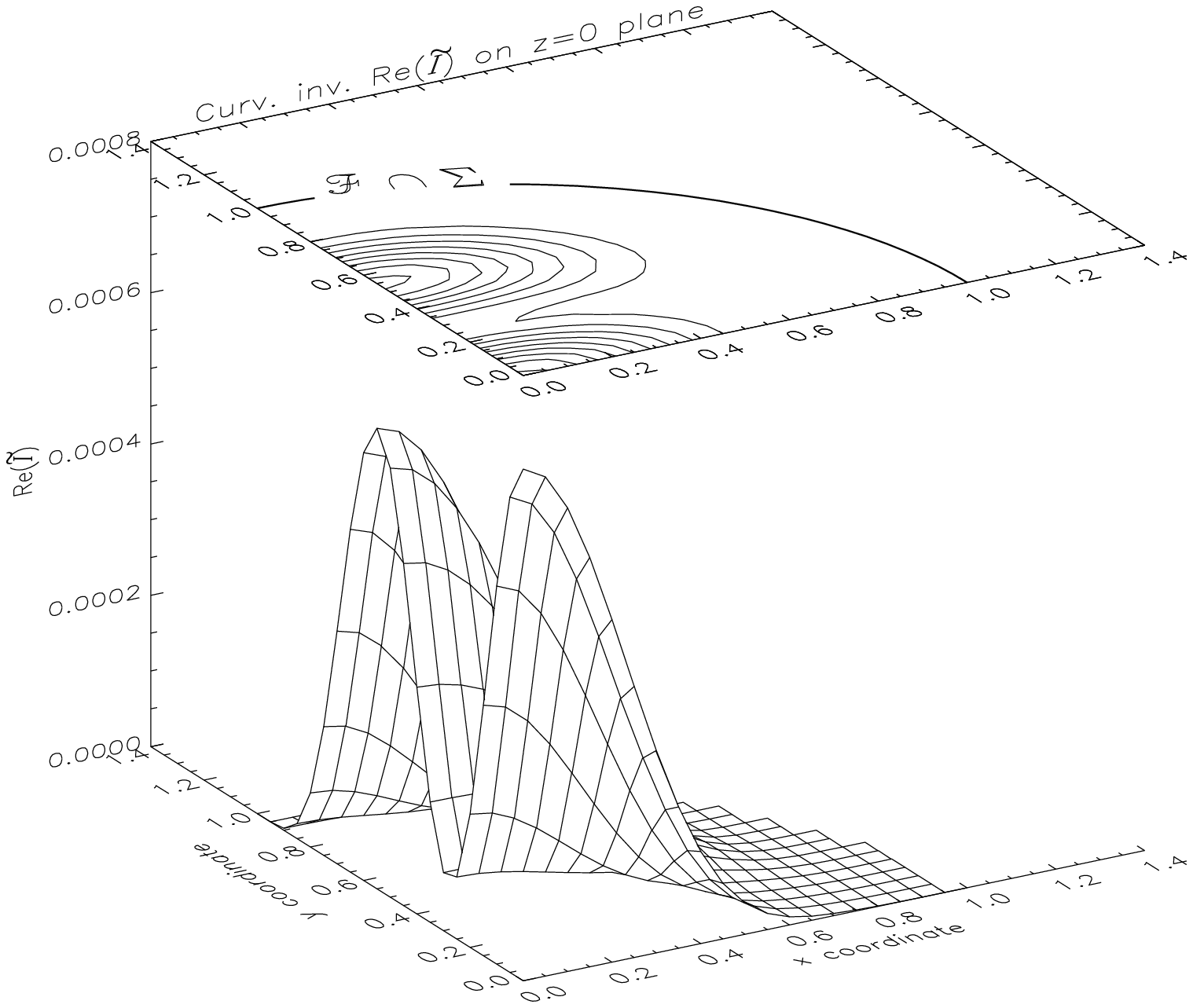}%
      \vskip0.5em
      \caption{\label{Anfangsdaten}The initial values for the metric
        component $h_{\underline{x}\underline{x}}$ (left) and the
        real part of the curvature invariant $\tilde{I}$ (right) on
        the $z=0$ plane. Only the physical part of the grid is shown.}
    \end{minipage}
  \end{center}
\end{figure}

\hspace{-\parindent}%
To calculate permissable initial data we use the numerical scheme
described in~\cite{Hu00nc}, to which we refer for details.
There, on an initial hypersurface we give a boundary defining function
$\bar\Omega$, which will be related to the conformal factor $\Omega$
and the solution $\phi$ of equation~(\ref{Yamabe}) by
$\Omega=\bar\Omega/\phi$, and a conformal 3-metric $h_{ab}$.
The 3-metric is chosen such that the tracefree part of the
extrinsic curvature of the surface $\bar\Omega=0$, the initial cut
${\cal S}$ of null infinity, must vanish.
\\
Then we solve the Yamabe equation,
\begin{eqnarray}
\label{Yamabe}
  & &
  4 \, \bar\Omega^2 \DeIII \phi
  - 4 \, \bar\Omega (\!\DIII^a \bar\Omega)(\!\DIII_a \phi) \nonumber\\
  & & \qquad
  - \left( \frac{1}{2} \RIII \, \bar\Omega^2 + 2 \bar\Omega \DeIII\bar\Omega
           - 3 (\!\DIII^a \bar\Omega) (\!\DIII_a \bar\Omega) 
    \right) \phi
  - \frac{1}{3} {\tilde k}^2 \phi^5
  \quad = \quad 0,
\end{eqnarray}
where $\DIII$, $\DeIII$, and $\RIII$ are the derivative operator, the
Laplace operator, and the Ricci scalar associated with $h_{ab}$, and
$\tilde k$ is a positive constant, which we choose to be $\tilde k =
3$.
\\
The solution $\phi$ of the Yamabe equation, the given free functions
$\bar\Omega$ and $h_{ab}$, and two more free functions, namely the trace
$k$ of the conformal extrinsic curvature of the initial slice and the Ricci
scalar $R$ of the conformal spacetime, define a set of data for the
conformal field equations.
\\
Our choices for the free functions are
\begin{subequations}
  \begin{eqnarray}
    \bar\Omega & = & \frac{1}{2} \left( 1 - \left(x^2+y^2+z^2\right) \right) \\
    h_{\underline{a}\underline{b}} & = &
      \left(\begin{array}{ccc}
        1 + \frac{1}{3} A \bar\Omega^2 \left( x^2 + 2 y^2 \right)
          & 0 & 0 \\
        0 & 1 & 0 \\
        0 & 0 & 1
      \end{array}\right) \\
    k & = & 0 \\
    R & = & 0.
  \end{eqnarray}
\end{subequations}
If $\phi$ is a solution of the Yamabe equation with respect to
$\bar\Omega$ and $h_{ab}$, then for any positive function $\theta$ the
solution of the Yamabe equation with respect to $\theta\bar\Omega$
and $h_{ab}$ is $\phi\sqrt{\theta}$.
Therefore, the calculated data depend on the location of
${\cal S}$ but not on the choice of $\bar\Omega$ in the interior, and
there is no loss of generality, if we choose $\bar\Omega$ to be
spherically symmetric.
\\
The 3-metric $h_{ab}$ is obtained by perturbing the $\underline{x}\underline{x}$
component away from Minkowski data.
The chosen perturbation has no obvious continuous symmetry, but it is 
reflection symmetric at the planes $x=0$, $y=0$, and $z=0$.
For that reason it is sufficient to plot the octant
$\{(x,y,z)|x\ge0,y\ge0,z\ge0\}$, although the calculation has been
performed for
$(x,y,z) \in [-1.25,1.25]\times[-1.25,1.25]\times[-1.25,1.25]$.
\\
The choices of $k$ and $R$ are pure gauge.
The function $k$ determines the initial time derivative $\Omega_0$ of
the conformal factor $\Omega$ through the relation (7)
of~\cite{Hu00nc}, the choice of $R$ determines
the values of certain curvature components $\RII_{ab}$ and $E_{ab}$ and
the variable $\omega$, obtained by applying the wave operator applied
to $\Omega$.
\\
To calculate initial data, we use a spectral bases of $46\times
44\times 44$ elements.
The maximum value of the constraint violation on the initial slice is
then of order $10^{-4}$.
\\
The left graphic in figure~\ref{Anfangsdaten} shows the metric component
$h_{\underline{x}\underline{x}}$ on the $z=0$ plane for a value of $A=1$.
\\
The plane $z=0$ is the plane, on which the perturbation of
$h_{\underline{x}\underline{x}}$ assumes its maximum in the physical
region, which is approximately $0.025$.
A priori, it is not clear, whether the perturbation is not just a
coordinate effect.
The right graph in figure~\ref{Anfangsdaten} shows the real part
$\Re(\tilde I)$ of the curvature invariant
\begin{eqnarray}
  \label{PhysCurvInv}
  \tilde I & = & \tilde \psi_{ABCD} \; \tilde \psi^{ABCD} \nonumber \\
           & = & \Omega^6 
                 \left(  E^{ab} E_{ab} - B^{ab} B_{ab} 
                         + 2 \, i \, E^{ab} B_{ab} \right),
\end{eqnarray}
where $\tilde \psi_{ABCD}$ is the Weyl spinor of the physical
spacetime, $E_{ab}$ the electric part of the rescaled conformal Weyl
tensor, $B_{ab}$ its magnetic part, and $i=\sqrt{-1}$.
Since this curvature invariant does not vanish, our data do not
correspond to Minkowski space.
\\
The curvature invariant $\Re(\tilde I)$ assumes its maximum value of
approximately $0.00068$ at the origin.
The maximum value for an amplitude $A=2.0$ is approximately $0.00270$.
The curvature invariant $\Re(\tilde I)$ is therefore approximately
proportional to $A^2$.
There is another local maximum at $y\approx 0.55$ with a value of
approximately $0.00061$. 
The physical length of the $y$ coordinate line connecting this maximum
with the origin is about $1.25$.
%
%
%
%
%

%% file: Zeitentwicklung.tex
%
%
%
\section{The time evolution}
\label{Zeitentwicklungsabschnitt}
\hspace{-\parindent}%
We split the discussion of the time evolution of our data into three
parts. 
The first part describes the conformal structure of the spacetime.
In the second part we describe how we proceed to reconstruct
properties of the associated physical spacetime, i.~e.\ the
proper time of observers.
Issues related to gravitational radiation, such as the
longtime decay of the Bondi mass, are dealt with in the third part.
\\
To evolve our data, we have to choose five gauge source functions, namely
$q=\ln(N/\sqrt{h})$, where $N$ is the lapse and $h$ the determinant of
$h_{ab}$, the three components of the shift $N^a$, and the Ricci
scalar $R$.
In our case the simplest choice,
\begin{subequations}
\begin{eqnarray}
  q   & = & 0 \\
  N^a & = & 0 \\
  R   & = & 0
\end{eqnarray}
\end{subequations}
is sufficient to cover the entire future of the initial data.
\\
The size of the next numerical time step is calculated from the
Courant-Friedrich-Levy condition at each step as described
in~\cite{Fr98ntb}.
Corresponding time slices of runs with different resolutions do in
general not coincide, since the size of the time step depends on the
3-metric $h_{ab}$, which is a variable of our system.
\\
Simultaneous to the time evolution equation we solve 15 ordinary
differential equations describing geodesics of the physical metric
$\tilde g_{ab}$ (cf.\ \cite{HuWXXxx} for numerical details).
These world lines represent observers in the physical spacetime.
Initially the observers are placed at coordinate values of $(0,0,0)$,
$(\pm\frac{1}{2},0,0)$, $(0,\pm\frac{1}{2},0)$,
$(0,0,\pm\frac{1}{2})$,
and $(\frac{\pm 1}{2\sqrt{3}},\frac{\pm 1}{2\sqrt{3}},\frac{\pm
  1}{2\sqrt{3}})$.
As initial tangent vector we choose the normal of the initial slice
and normalise it with respect to the physical metric $\tilde g_{ab}$.
\\
In addition to the observers in physical spacetime we calculate the
orbits of 1986 ``Bondi observers moving along generators of null
infinity''.
Their orbits define a discretisation of \Scri{} and enable us to
calculate radiative quantities (cf.\ \cite{HuWXXxx} for details).
The Bondi observers are placed at a uniform grid parametrising the
initial cut ${\cal S}$ of \Scri{}.
They are placed at the north and the south pole and at
$64\times31$ gridpoints covering $(\vartheta,\varphi) \in
]0,\pi[\times[0,2\pi[$.
\\
An important property of every numerical simulation is an error
estimate.
The convergence analysis of runs with spatial grids of $(50^3)$,
$(100^3)$, and $(150^3)$ gridpoints performed on slices at $t=0.0$,
$0.24$, $0.49$, $0.72$, and $0.92$ gives an estimate for the absolute
maximal error in any variable of less than $0.001$.
This value agrees with the errors obtained when reproducing exact
solutions with the same code~\cite{Hu99as}.
We have also performed convergence analyses for each plot shown.
If not explicitely mentioned, the error estimate of the convergence
analysis suggests, that the error is at most of the order of the line
thickness in the plots.
\subsection{The conformal spacetime}
\begin{figure}[htbp]
  \begin{center}
    \begin{minipage}[t]{8cm}
      \vskip2em      
      \includegraphics[width=8cm]{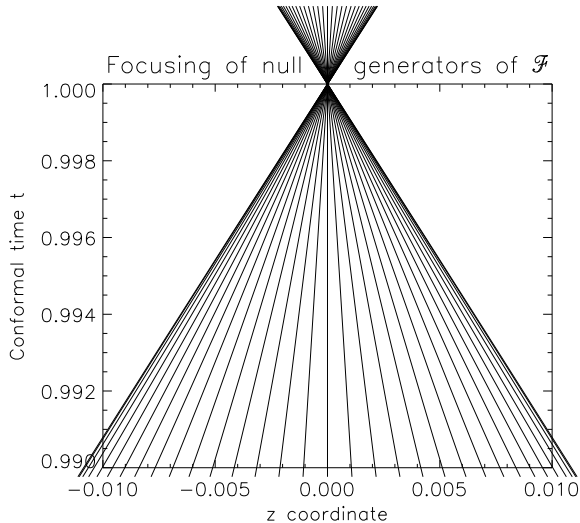}%
      \vskip0.5em
      \caption{\label{FaserBuendel}The projection of representative null
        generators of \Scri{} onto the $(z,t)$ plane near the 
        focal point.}
    \end{minipage}
  \end{center}
\end{figure}

\begin{figure}[htbp]
  \begin{center}
    \begin{minipage}[t]{8cm}
      \includegraphics[width=8cm]{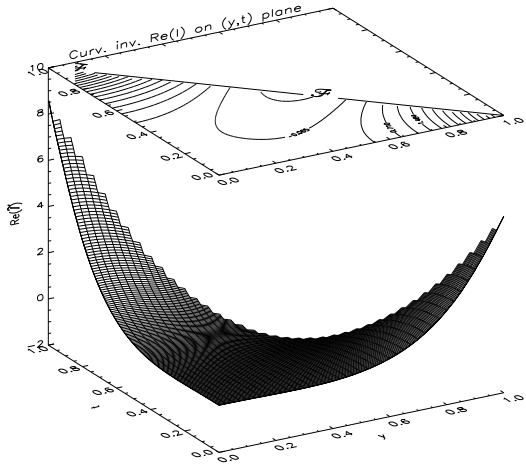}%
      \vskip0.5em
      \caption{\label{ReConfIy}Time evolution of real part of
        conformal curvature invariant $I$ on the positive $y$ axis (only
        the physical region of conformal spacetime is shown).}
    \end{minipage}
  \end{center}
\end{figure}

\hspace{-\parindent}%
During the evolution of our data we monitor in particular the
behaviour of the conformal factor~$\Omega$.
We will have covered the entire physical future of the initial data,
if the region $\{\Omega > 0\}$ vanishes for one slice, since the
physical spacetime is identical with the set $\{\Omega > 0\}$.
The orbits of the Bondi observers generate the surface
$\{\Omega=0\}$.
In figure~\ref{FaserBuendel}
we plot their orbits near their focal point.
\\
Due to practical reasons we restrict ourselves to the projection onto
the plane $\{x=0,y=0\}$ and a representative selection 
of our 1986 Bondi observers, namely the observers initially
placed at the 33 points $(\vartheta,\varphi) = (0,0), (\pi/32,0),
\ldots, (\pi,0)$.
Plots showing the projections to the planes $\{y=0,z=0\}$ and 
$\{x=0,z=0\}$ and projections of all the other orbits give
corresponding results.
\\
It should also be pointed out that the size of a grid cell in the
$150^3$ run we present is $\Delta t \times \Delta x \times \Delta y
\times \Delta z \approx  0.003 \times 0.017^3$ near the focal
point.
The generators of \Scri{} clearly meet within the volume of one
grid cell.
The result from a $100^3$ run is visibly indistinguishable from the
$150^3$ run, in the $50^3$ run the focal point has a slightly
smaller $t$ coordinate, namely $\approx 0.996$, which is still in
excellent agreement, since the size of the $50^3$ grid cells is
$\approx  0.009 \times 0.05^3$ near the focal point, which is
larger than the deviation of the runs.
\\
We have continued our calculation beyond the focal point up to
$t=1.1$ to check regularity of the conformal spacetime at the focal
point.
Since we have completely covered the physical future of our initial
slice at $t=1$ already, it makes no sense to integrate further.
Even beyond the focal point the conformal spacetime stays regular.
Therefore the focal point is an excellent candidate for a regular
$i^+$.
And indeed, as we will see later, physical observer will reach this
point after infinite proper time.
\\
Since we have found a complete $\Scri{}^+$ and a regular $i^+$, the
constructed spacetime possesses qualitatively the same asymptotic
structure as the Minkowski space.
Quantitatively, there are differences in the asymptotics as can be seen
from figure~\ref{ReConfIy},
where we plot the real part $\Re(I)$ of the conformal curvature invariant 
\begin{eqnarray}
  \label{ConformalCurvInv}
  I & = & \psi_{ABCD} \; \psi^{ABCD} \nonumber \\
    & = & E^{ab} E_{ab} - B^{ab} B_{ab} + 2 \, i \, E^{ab} B_{ab}
\end{eqnarray}
on the physical portion of the grid.
The quantity $\psi_{ABCD}$ denotes the rescaled Weyl spinor.
\\
There is a significant amount of rescaled conformal curvature at
$\Scri{}^+$ and at $i^+$, indicating a non-vanishing fall-off
coefficient in the expansion of the corresponding physical curvature
invariant $\Re(\tilde I)$ at infinity.
\subsection{Reconstructing the associated physical spacetime}
\begin{figure}[htbp]
  \begin{center}
    \begin{minipage}[t]{8cm}
      \includegraphics[width=8cm]{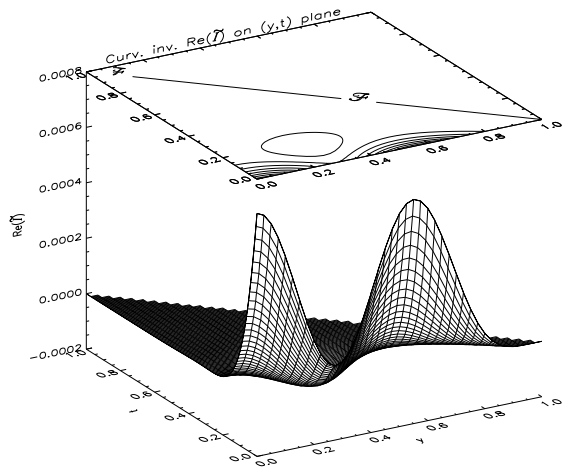}%
      \vskip0.5em
      \caption{\label{ReIy}Time evolution of the real part of the physical
        curvature invariant $\tilde I$ on the positive $y$ axis.}
    \end{minipage}
  \end{center}
\end{figure}

\begin{figure}[htbp]
  \begin{center}
    \begin{minipage}[t]{8cm}
      \includegraphics[width=8cm]{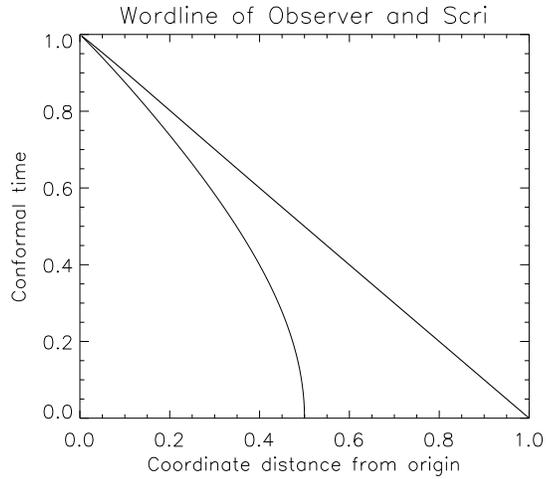}%
      \vskip0.5em
      \caption{\label{BeobachterWeltlinie}The thin line represents the
        world line of an observer (timelike physical geodesic) starting at 
        $(x,y,z) =
        (\frac{1}{2\sqrt{3}},\frac{1}{2\sqrt{3}},\frac{1}{2\sqrt{3}})$
        and going to $i^+$.
        The thick line represents $\Scri{}^+$.
        The particular behaviour of the world line near $i^+$ is due
        to the particular choice of its initial velocity.}
    \end{minipage}
  \end{center}
\end{figure}

\begin{figure}[htbp]
  \begin{center}
    \begin{minipage}[t]{8cm}
      \includegraphics[width=8cm]{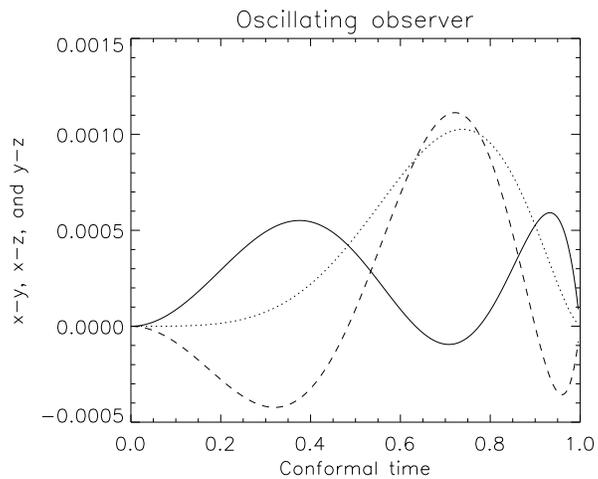}%
      \vskip0.5em
      \caption{\label{BeoZittern}Oscillation of an observer in coordinate
        space: The differences $x-y$ (solid), $x-z$ (dashed), and
        $y-z$ (dotted) as a function of conformal time $t$ which 
        would all be $0$ for Minkowski data ($A=0$).}
    \end{minipage}
  \end{center}
\end{figure}

\begin{figure}[htbp]
  \begin{center}
    \begin{minipage}[t]{16cm}
      \includegraphics[width=7.75cm]{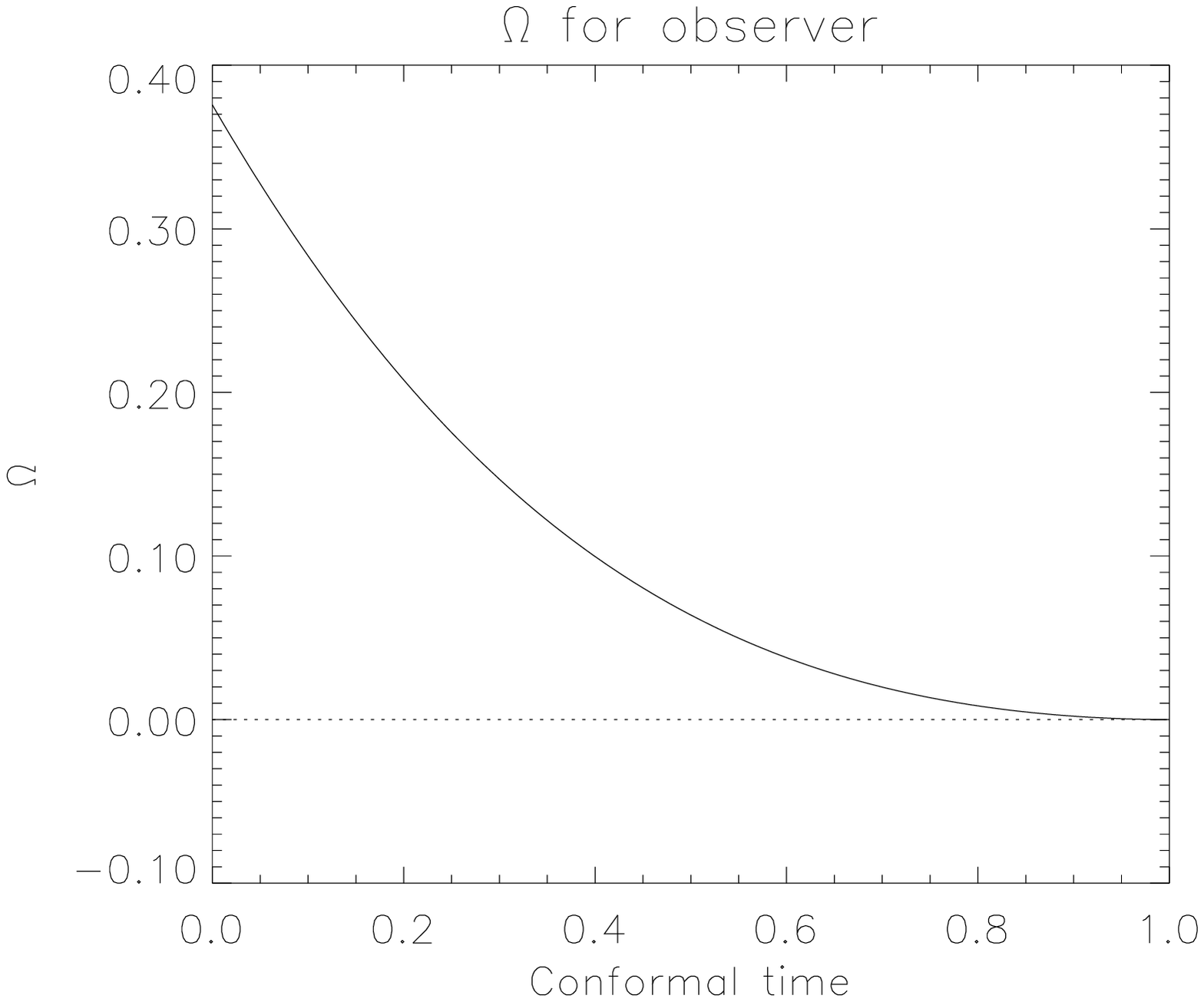}%
      \hspace{.5cm}
      \includegraphics[width=7.75cm]{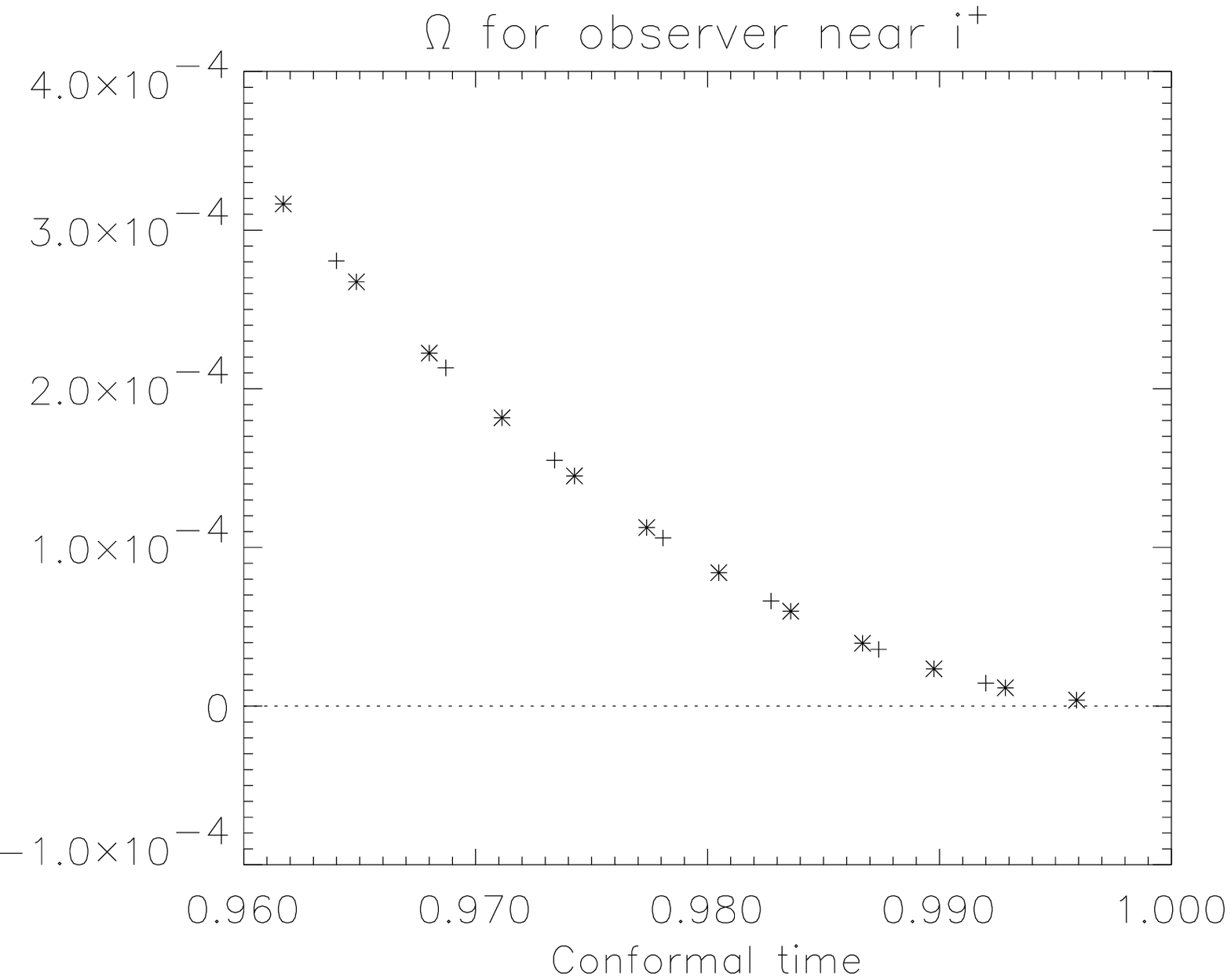}%
      \vskip0.5em
      \caption{\label{NaheiPlus}Left: Conformal factor $\Omega$ along
        an observer world line for the entire time evolution. Right:
        The same near $i^+$ (right) for a $50^3$~($\times$), an
        $100^3$~($+$), and an $150^3$~($\star$) run.}
    \end{minipage}
  \end{center}
\end{figure}

\begin{figure}[htbp]
  \begin{center}
    \begin{minipage}[t]{8cm}
      \includegraphics[width=8cm]{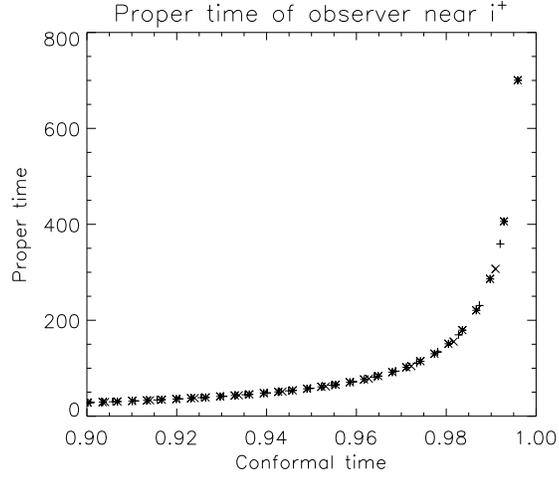}%
      \vskip0.5em
      \caption{\label{ProperNaheiPlus}Proper time along an
        observer world line near $i^+$ for a $50^3$~($+$), an
        $100^3$~($\times$), and an $150^3$~($\star$) run.}
    \end{minipage}
  \end{center}
\end{figure}

\begin{figure}[htbp]
  \begin{center}
    \begin{minipage}[t]{8cm}
      \includegraphics[width=8cm]{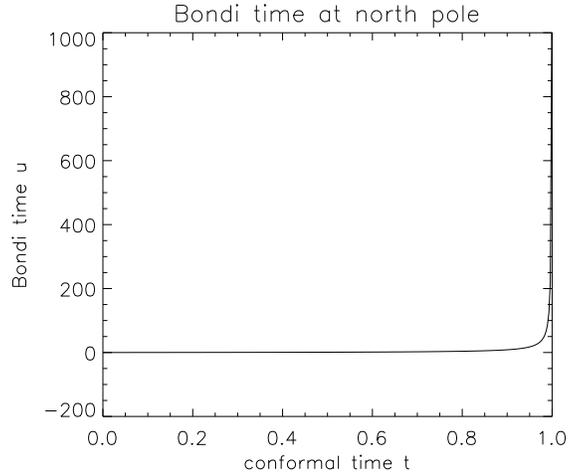}%
      \vskip0.5em
      \caption{\label{BonditatNP}Bondi time of the Bondi observer at
        the north pole as a function of conformal time.}
    \end{minipage}
  \end{center}
\end{figure}

\hspace{-\parindent}%
When we reconstruct the physical spacetime from the conformal
spacetime, we have to consider two types of quantities.
\\
The first type consists of those quantities which can be expressed as
a regular expression in the variables of the conformal field equations.
Figure~\ref{ReIy}
shows the time evolution of the value of $\Re(\tilde I)$, which is
such a quantity, on the $y$ axis.
We see that the time evolution of the two maxima of the initial data is
dominated by a rapid decay towards future null and future timelike
infinity, although there is another, smaller extremum forming in
between the initial maxima.
This smaller extremum can best be seen in the contour plot on top of
the surface plot.
In the contour plot we have also marked the location of $\Scri{}^+$.
\\
With a code performing the numerical integration in physical spacetime
we would only be able to calculate figure~\ref{ReIy}, but not
figure~\ref{ReConfIy}.
Since the physical curvature invariant decays so rapidly to zero (at
a conformal time of $0.6$ it has already decayed by a factor of order
$10000$), only a very accurate physical code could resolve the
fall-off at a large physical time.
In the conformal picture the decay has been factored out by choosing
the rescaled conformal Weyl tensor as variable.
We calculate the conformal invariant $\Re(I)$ and the conformal factor
$\Omega$, which do not change dramatically during the whole time
evolution, and then get the physical invariant by multiplying the
conformal curvature invariant with the appropriate powers of the
conformal factor (equations \ref{PhysCurvInv} and
\ref{ConformalCurvInv}).
\\
A convergence analysis shows that the numerical prediction of the
behaviour of $\Re(\tilde I)$ for a $150^3$ calculation is
indistinguishable from the $100^3$ calculation for the whole
calculation up to $i^+$.
At the last slice of the $100^3$ run before $i^+$  the curvature
invariant has decayed by a factor of more than $10^{20}$.
Due to the effect of rounding errors, a physical code working
with 8~byte reals could not resolve the decay over such a large range,
regardless of the grid size required to achieve the required
accuracy.
\\
The physical metric $\tilde g_{ab}= \Omega^{-2} g_{ab}$ blows up at
infinity, since the conformal metric $g_{ab}$ is regular everywhere.
The second type of quantity consists of those quantities which
describe physical distances.
They, of course, must blow up when approaching $i^+$ or $\Scri{}^+$.
\\
A typical example is the proper time of observers.
We will see in the following paragraphs, that the question ``For how
long can we numerically calculate the measurements of an observer?'' is
closely tied to the numerical question ``How well can we resolve the
neighbourhood of the surface $\{\Omega=0\}$?''.
\\
Figure~\ref{BeobachterWeltlinie} 
shows the world line of a representative observer, who is initially
placed at $(x,y,z) =
(\frac{1}{2\sqrt{3}},\frac{1}{2\sqrt{3}},\frac{1}{2\sqrt{3}})$, in an
$(r=\sqrt{x^2+y^2+z^2},t)$ plot.
The world line runs into $i^+$ as expected.
\\
We can plot the $x$, $y$, and $z$ coordinates of the world line of the
observer as a function of conformal time.
If our spacetime were a representation of Minkowski space, and if we
were to choose the same gauge source functions, the differences $x-y$, $x-z$,
and $y-z$ would vanish for all times, due to symmetry reasons.
Figure~\ref{BeoZittern}
shows, that this is not the case in the computed spacetime.
The observer oscillates around the Minkowski orbit --- he is pushed
around by the gravitational wave.
The amplitude of the oscillation is small compared to the coordinate
values. 
By plotting the difference we make the oscillation visible.
\\
A plot of the conformal factor along the world line of our observer gives
the left graph of figure~\ref{NaheiPlus}.
The right graph of that figure shows the vicinity of $i^+$ for the
$50^3$ ($\times$), the $100^3$ ($+$), and the $150^3$ ($\star$) runs.
Obviously, the finer the resolution the closer we get to the
(quadratic) zero of $\Omega$.
\\
When calculating the physical geodesics we also calculate the proper
time $\tau$ as a function of the conformal time $t$.
In Figure~\ref{ProperNaheiPlus}
we show the result for our observer near $i^+$.
In the $50^3$ run we have covered a proper time interval of
$[0,300]$, in the $100^3$ run an interval of $[0,350]$, and in the
$150^3$ run an interval of $[0,700]$.
When the integration time is measured in units of the
(Bondi) mass $m_B(0)$ on the initial slice, the last
number is more than $600\,000 \; m_B(0)$.
\\
In figure~\ref{BonditatNP}
we plot the Bondi time $u$  of an observer initially placed at the
north pole of \Scri{} for the conformal time interval $[0,1]$.
Most conspicuous is the rapid growth near $i^+$, with the
largest value $u \approx 950$.
\subsection{The decay of the Bondi mass}
\begin{figure}[htbp]
  \begin{center}
    \begin{minipage}[t]{8cm}
      \includegraphics[width=8cm]{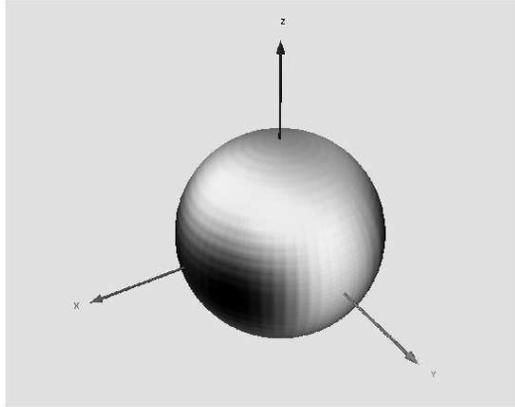}%
      \vskip0.5em
      \caption{\label{QuadrupolMoment}$\psi_2$ on the initial cut of
        \Scri{}. 
        Values range from $-0.148$ (black) to $0.115$ (white).}
    \end{minipage}
  \end{center}
\end{figure}

\begin{figure}[htbp]
  \begin{center}
    \begin{minipage}[t]{8cm}
      \includegraphics[width=8cm]{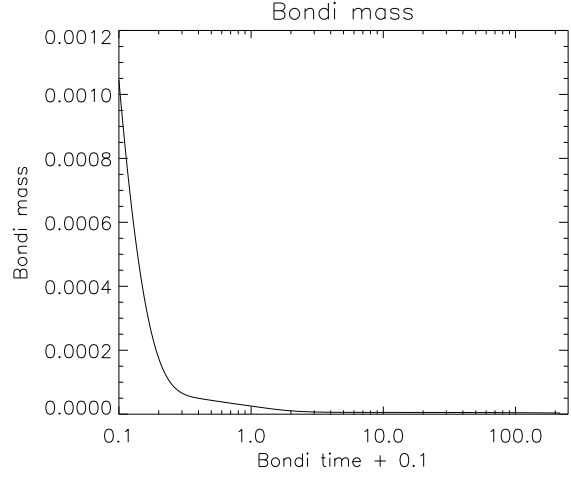}%
      \vskip0.5em
      \caption{\label{BondiMasseBondizeit}Bondi mass as a function
        of Bondi time.}
    \end{minipage}
  \end{center}
\end{figure}

\begin{figure}[htbp]
  \begin{center}
    \begin{minipage}[t]{8cm}
      \includegraphics[width=8cm]{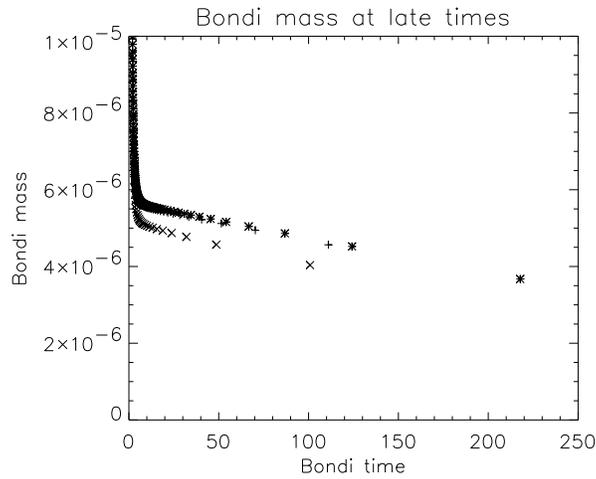}%
      \vskip0.5em
      \caption{\label{BondiMassSpaet}Bondi mass as a function
        of Bondi time at late times for $50^3$ ($\times$), $100^3$
        ($+$), and $150^3$ ($\star$) runs.}
    \end{minipage}
  \end{center}
\end{figure}

\hspace{-\parindent}%
Here we shortly sketch how we calculate the Bondi mass.
Details are given in~\cite{HuWXXxx}.
\\
The Bondi mass of a cut ${\cal S}_u$ of \Scri{} is given by
\begin{equation}
  \label{BondiMasse}
  m_B = - \int_{{\cal S}_u} \left( \check\psi_2 
            + \check\sigma \dot{\bar{\check\sigma}} \right) \; d^2{\cal S},
\end{equation}
where $\check{\hphantom{g}}$ denotes quantities with respect to a conformal
metric $\check g_{ab} = \alpha^{-2} g_{ab}$ in which \Scri{} is
expansion free.
The quantity $\check\sigma$ is one of the spin coefficients in the
Newman-Penrose formalism, its Bondi time derivative
$\dot{\check\sigma}$ is the news function.
\\
When calculating the world lines of Bondi observers, we keep track of
the evolution of $\alpha$ for each observer and we
parallelly transport with respect to $\check g_{ab}$ the null frame
associated with the observer.
\\
Since the $u=\const$ cuts do in general not coincide with the
$t=\const$ cuts, we have to locate the $u=\const$ cuts and interpolate
onto them from the $t=\const$ cuts before we perform the integration.
\\
Once the initial value of the Bondi mass $m_B(0)$ is determined, there
is an alternative calculation of the Bondi mass, using the mass loss formula:
\begin{equation}
  \label{BondiMasseAlt}
  m_B(u) = m_B(0) - \int_0^u \int_{{\cal S}_{u'}} 
                   \left( \dot{\check\sigma} \dot{\bar{\check\sigma}} 
                   \; d^2{\cal S} \right) \; du'.
\end{equation}
The latter method usually gives much smaller errors.
Hence we use~(\ref{BondiMasseAlt}) to calculate the time evolution of
the Bondi mass.
The initial value is taken from the most accurate calculation, namely
the $150^3$ run.
\\
In our setup $\check\sigma$ is $0$ on the initial slice.
Figure~\ref{QuadrupolMoment}
shows the integrand $\check\psi_2$ on the initial cut of \Scri{}.
Obviously there is a strong high order moment.
Since the maximal value of $|\check\psi_2|$ is about $0.15$, and the
initial value of the Bondi mass is about $0.00104$, the high order
moments are by a factor of order $100$ stronger than the monopole
moment, which is the Bondi mass.
\\
Figure~\ref{BondiMasseBondizeit}
shows the Bondi mass as a function of the Bondi time $u$ on a logarithmic
scale (the abscissa is $\log(u+0.1)$).
We observe a rapid decay in a first stage which lasts until $u\approx
0.15$.
Then there is a slower decay lasting until $u\approx 1.1$.
\\
To see what happens after this stage we change the scale of the
ordinate in figure~\ref{BondiMassSpaet},
where we plot the Bondi mass against a non-logarithmic time scale.
\\
Since the Bondi mass has already decayed by a factor of about $200$,
even our small numerical error becomes an issue.
The result from the $50^3$ (indicated by $\times$) run significantly
differs from the results for the $100^3$ ($+$) run, whereas the later
almost coincides with the results of the $150^3$ ($\star$) run.
\\
Analytically the Bondi mass of a spacetime with a regular $i^+$ must
vanish at $i^+$.
It is not clear, whether the final value of the computed Bondi mass would
eventually vanish, if we integrated even longer, or whether this
offset is due to a numerical error in the initial value for the Bondi
mass, which mainly depends on how accurate the provided initial data
solve the constraints.
%
%
%
%

%% file: Zusammenfassung.tex
%
%
%
\section{Conclusion}
\hskip-\parindent{}%
We have calculated the future time evolution of hyperboloidal
gravitational wave data, which do not possess any continuous
symmetry.
We have seen that the conformal approach allows us to cover the
entire physical future of these data with a finite grid and to
determine the decay of curvature invariants over ranges unreachable by
codes working in physical spacetime.
\\
Due to the use of higher order methods we can use fairly coarse grids.
Since the grids used are coarse, we only need moderate amounts of
computer resources, in particular our time evolution on a
Origin2000 with R10000 processors requires 
\begin{itemize}
\item less than 15 minutes on 8 processors for a $50^3$ run, where we need
  87~time steps to cover $i^+$,
\item less than 2 hours on 16 processors for a $100^3$ run, where we  need
  173~time steps to cover $i^+$, and 
\item less than 6 hours on 27 processors for a $150^3$ run, where we
  need 263 ~time steps to cover $i^+$.
\end{itemize}
Already in the $50^3$ run the error is at most a few percent.
With the $150^3$ run we achieve an error of less than one part in
thousand.
\section*{Acknowledgement}
\hskip-\parindent{}%
I would like to thank H.~Friedrich and B.~Schmidt for their help and
support and J.~Winicour for very useful comments on the manuscript.
\\
I acknowledge the use of program code for the determination of the
Bondi observers which has been written in collaboration with
M.~Weaver.
\\
W.~Benger has produced figure~\ref{QuadrupolMoment} for me.
I thank him for that favour.
\\
Last but not least I acknowledge K.~St\"uben from the Gesellschaft f\"ur
Mathematik und Datenverarbeitung, who put the Algebraic Multigrid
Library AMG at my disposal, and M.~Frigo and S.~G.~Johnson who wrote FFTW
and made it publically available for the scientific community.
Both libraries are used when calculating initial data.
%
%
%

%% file: biblio.tex
\bibliography{biblio}
\bibliographystyle{prsty}